\documentclass[twocolumn,aps,prl,showpacs]{revtex4}

\usepackage{amsmath}
\usepackage{epsfig}
\usepackage{graphicx}

\begin{document}

\title{Divergent four-point dynamic density correlation function of  
a glassy colloidal suspension: a diagrammatic approach}
\author{Grzegorz Szamel}
\affiliation{Department of Chemistry, 
Colorado State University, Fort Collins, CO 80523}

\date{\today}


\begin{abstract}
We use a recently derived diagrammatic formulation of the dynamics of
interacting Brownian particles [G. Szamel, J. Chem. Phys. \textbf{127}, 084515 (2007)] 
to study a four-point dynamic density correlation function. We re-sum 
a class of diagrams which separate into two disconnected components 
upon cutting a single propagator. 
The resulting formula for the four-point correlation function
can be expressed in terms of three-point functions closely related 
to the three-point susceptibility 
introduced by Biroli \textit{et al.} [Phys. Rev. Lett. \textbf{97},
195701 (2006)] and the standard two-point correlation function. The four-point 
function has a structure very similar to that proposed by Berthier and collaborators
[Science \textbf{310}, 1797 (2005), J. Chem. Phys. \textbf{126}, 184503 (2007)]. 
It exhibits a small wave vector divergence at the mode-coupling transition.
\end{abstract}
\maketitle

\textit{Introduction} --- 
It has become clear over the past decade that upon approaching the glass
transition, the liquid's dynamics not only get slower but
also become increasingly heterogeneous \cite{reviews}. 
Naturally, one of the very first questions that arose after the discovery
of dynamic heterogeneities was that of their spatial extent. Indeed, some of the
very first simulational studies of dynamic heterogeneities tried to estimate
their size and showed that it increases upon cooling \cite{KobDonati}.
The discovery of growing dynamic heterogeneities posed a challenging problem for 
the mode-coupling theory \cite{Goetze,Shankar} of the glass
transition. First, according to the standard formulation 
of this theory \cite{Goetze}, the glass transition 
is supposed to be a small spatial scale 
phenomenon arising from a self-consistent caging of individual particles
in their first solvation shells. Second, the mode-coupling theory 
relies upon a factorization approximation for a four-point dynamic density
correlation function. Thus, it cannot even be expected to 
describe highly non-trivial spatiotemporal dependence of various 
four-point 
correlation functions that were introduced to monitor dynamic heterogeneities.

The above described problem was first addressed by Biroli and Bouchaud (BB) \cite{BB}. 
Inspired by an earlier study \cite{Franz} 
of the so-called schematic mode-coupling equations, BB argued, using
a field theoretical formulation of many-particle dynamics, that the 
mode-coupling theory should be understood as a saddle point approximation derived from 
an action functional. Furthermore, they argued 
that a four-point dynamic density correlation function
could be calculated by inverting a second functional derivative of the same
functional and that this procedure 
corresponds to a re-summation of ladder diagrams. They analyzed the
convergence of the diagrammatic series and showed that their 
four-point function diverges at the transition point of the mode-coupling theory.
In a very interesting later contribution Biroli, Bouchaud, Miyazaki and Reichman 
(BBMR) \cite{BBMR} used a more traditional, projection operator based \cite{Goetze}
version of the mode-coupling theory 
and showed that the matrix which determines the convergence of BB's four-point function
is exactly the same as the matrix which describes long wavelength
properties of a certain three-point dynamic susceptibility describing
the response of the intermediate scattering function to an external potential.
Finally, in a pair of recent papers \cite{JCPa,JCPb} 
Berthier \textit{et al.} 
used a field theoretical approach to argue that the most divergent part of the 
four-point dynamic density correlation function is given by, roughly speaking, a 
product of two three-point dynamic susceptibilities. According to Ref. \cite{JCPa}
these three-point susceptibilities can be expressed in terms of ladder diagrams
(in qualitative agreement with BBMR)
and thus the most divergent part of the four-point function
is represented by, roughly speaking, the sum of ``squared ladder'' 
diagrams \cite{commentsubd}. 

The goal of this Letter is to analyze a four-point dynamic density correlation 
function of a glassy colloidal suspension. The reason for considering 
a colloidal system is twofold. First, experiments provide a wealth of information
about the motion of individual colloidal particles and thus enable detailed tests
of various theoretical predictions. Second, the simplest model of a colloidal
suspension, a system of interacting Brownian particles, is technically simpler to
study than a simple fluid. To analyze the four-point function we will use a recently
derived \cite{GSdiagram} 
diagrammatic formulation of the dynamics of interacting Brownian particles. 
This formulation can use the same tools, \textit{e.g.} 
re-summations of classes of diagrams, as a field theory based approach.
It is, however, closer to the standard, projection operator
based derivation of the mode-coupling theory. 

In contrast to BB, we focus on singly connected diagrams, \textit{i.e.}
diagrams which separate into two disconnected components upon cutting
through a single propagator. We will show that a sum of such diagrams
results in a contribution to the four-point correlation function that diverges at the 
transition point of the mode-coupling theory. The structure of this 
contribution is very similar to the structure of the most divergent part
of the four-point function obtained by Berthier \textit{et al.} \cite{JCPa,BerthierS}. 

Since the derivation of the main result is rather tedious, we will first
define our four-point function and present the main result.
We will outline the derivation of this result in the latter part of this Letter.

\textit{Four-point correlation function} --- A number of different 
functions have been introduced
to monitor dynamic heterogeneities \cite{different4pf}.
In particular, two recent studies  \cite{FS1,CGJMP,BerthierPRE} used
$N\left<e^{-i\mathbf{k}\cdot(\mathbf{r}_1(t)-\mathbf{r}_1)}
e^{i\mathbf{k}\cdot(\mathbf{r}_2(t)-\mathbf{r}_2)}
\delta(\mathbf{r}-\mathbf{r}_{12})\right>$, where $\mathbf{r}_i(t)$ denotes
the position of particle $i$ at time $t$, $\mathbf{r}_i(0)\equiv \mathbf{r}_i$, 
$\mathbf{r}_{12}=\mathbf{r}_1-\mathbf{r}_2$, and 
$N$ is the number of particles. 
This function quantifies 
correlations between evolutions of two particles that were initially separated
by $|\mathbf{r}|$. Its Fourier transform with respect to
$\mathbf{r}$ can be written as 
$N\left<e^{-i\mathbf{k}\cdot(\mathbf{r}_1(t)-\mathbf{r}_2(t))}
e^{i(\mathbf{k}-\mathbf{q})\cdot\mathbf{r}_{12}}\right>$.
In this work, we consider a collective version of the latter expression,
\begin{equation}\label{S4}
S_4(\mathbf{k};\mathbf{q};t) = \frac{1}{N}
\left<n_2(\mathbf{k},-\mathbf{k};t)n_2^*(\mathbf{k}-\mathbf{q},-\mathbf{k}+\mathbf{q})
\right>,
\end{equation} 
where $n_2(\mathbf{k}_1,\mathbf{k}_2;t)$ denotes 
a projected two-particle density \cite{Fourier},
\begin{equation}\label{n2def}
n_2(\mathbf{k}_1,\mathbf{k}_2;t) = \left( 1 - \mathcal{P}_0 - \mathcal{P}_1 \right)
\sum_{i,j}e^{-i\mathbf{k}_1\cdot\mathbf{r}_{i}(t)
-i\mathbf{k}_2\cdot\mathbf{r}_{j}(t)}.
\end{equation}
In Eq. (\ref{n2def}), $\mathcal{P}_0$ and $\mathcal{P}_1$ are projection operators onto
the zero-particle and one-particle density, respectively \cite{GSdiagram}. 
In the following, we will often use 
an abbreviated notation and, \textit{e.g.}, 
write $n_2(\mathbf{k}_1,\mathbf{k}_2;t)$ as $n_2(1,2;t)$ or as $n_2(t)$;
in addition, $n_2(t=0) \equiv n_2$.

The projection operators in Eq. (\ref{n2def}) remove 
a non-zero average value and
a non-vanishing projection onto the one-particle density. 
The terms projected out 
do not contribute to the divergent part of $S_4$. 
Moreover, consistent usage of projected many-particle densities 
analogous to $n_2$  
has important technical advantages \cite{GSdiagram,Andersen}.  
In particular, bare inter-particle interactions
are automatically renormalized by equilibrium correlation functions.

\begin{figure}
\includegraphics[scale=.22]{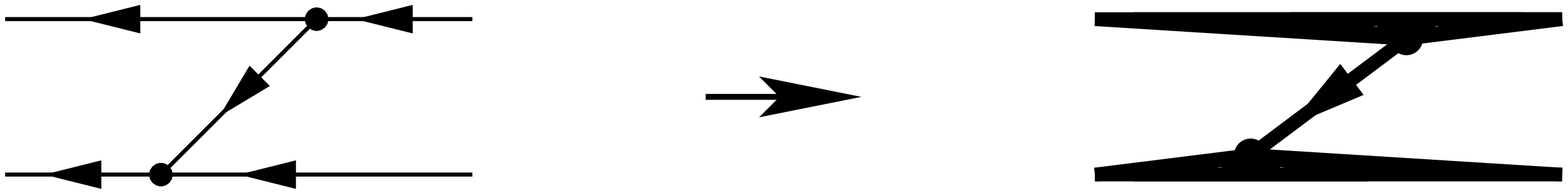}
\caption{Upon re-summation the simplest bare singly connected  
diagram on the left is replaced by a renormalized diagram on the right.} 
\label{f:diag0}
\end{figure}

\textit{Contribution to $S_4$ due to singly connected diagrams} --- 
In this work we focus on the 
diagrams which separate into two disconnected components upon cutting a single bond,
with each component containing
at least a single line of bonds from the initial time $0$ to the final time $t$
(the precise correspondence between $S_4$ and the diagrams will be detailed below).
The simplest diagram of this type is shown on the left in Fig. \ref{f:diag0}. 
In this diagram, 
bond \includegraphics[scale=.22]{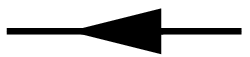} represents a bare propagator and 
vertices \includegraphics[scale=.22]{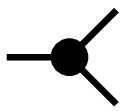}, 
and \includegraphics[scale=.22]{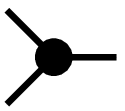}
represent the left and right vertices, 
respectively \cite{GSdiagram}. Upon a re-summation,
the bare singly connected diagram on the 
left in Fig. \ref{f:diag0} is replaced by the diagram on the right of this figure.  
The latter diagram involves two new functions,
closely related to the three-point susceptibility of BBMR, which are represented
by \includegraphics[scale=.22]{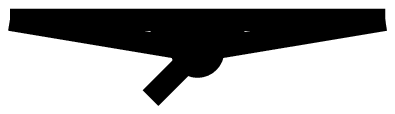} 
and \includegraphics[scale=.22]{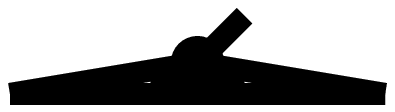}, and full propagator represented by
\includegraphics[scale=.22]{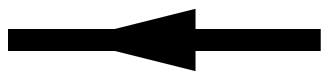}. 

The main result of this note is the following formula for the part of $S_4$
which becomes divergent at the mode-coupling transition; 
this formula results from an approximate re-summation indicated in Fig. \ref{f:diag0}:
\begin{eqnarray}\label{finalS4om}
S_4(\mathbf{k};\mathbf{q};t) &=&  \frac{2}{nS(q)} 
\int \frac{d\omega}{2\pi} 
\chi^2_{-\mathbf{q}}(-\mathbf{k};t;-\omega) \nonumber \\ &&  \times
G(q;\omega) \chi^1_{-\mathbf{q}}(\mathbf{k};t;\omega)
\end{eqnarray}
Here $G(q;\omega)$ is a Fourier transform of the full propagator, 
$G(q;\omega)=\int dt e^{i\omega t} G(q;t)$,  
and $G(k;t)$ is defined in terms
of the intermediate scattering function $F(k;t)$, $G(k;t) = \theta(t) F(k;t)/S(k)$.
Furthermore, in Eq. (\ref{finalS4om}) 
$\chi^1_{\mathbf{q}}(\mathbf{k};t;\omega)$ and 
$\chi^2_{\mathbf{q}}(\mathbf{k};t;\omega)$ are three-point functions. They are 
related by $\chi^2_{\mathbf{q}}(\mathbf{k};t;-\omega) = 
e^{-i\omega t} \chi^1_{\mathbf{q}}(\mathbf{k}-\mathbf{q};t;\omega)$.
Function $\chi^1_{\mathbf{q}}(\mathbf{k};t;\omega)$ 
satisfies the following equation of motion:
\begin{widetext}
\begin{eqnarray}\label{eomchi3}
\lefteqn{\int_{0}^t dt' \left(\delta(t-t')+M^{\mathrm{irr}}(k;t-t')\right)
\frac{\partial \chi^1_{\mathbf{q}}(\mathbf{k};t';\omega)}{\partial t'}  
+ \frac{D_0 k^2}{S(k)}\chi^1_{\mathbf{q}}(\mathbf{k};t;\omega) 
+ \int_{0}^{t} dt' \int\frac{d\mathbf{k}'}{(2\pi)^3}
\frac{n D_0 k}{|\mathbf{k}+\mathbf{q}|} 
} \nonumber \\ &&  \times
v_{\mathbf{k}}(\mathbf{k}',\mathbf{k}-\mathbf{k}')
\chi^1_{\mathbf{q}}(\mathbf{k}';t-t';\omega) F(|\mathbf{k}-\mathbf{k}'|;t-t')
v_{\mathbf{k}+\mathbf{q}}(\mathbf{k}'+\mathbf{q},\mathbf{k}-\mathbf{k}')
\frac{\partial F(|\mathbf{k}+\mathbf{q}|;t')}{\partial t'} = 
\mathcal{S}_{\mathbf{q}}(\mathbf{k};t;\omega)
\end{eqnarray}
\end{widetext}
In Eq. (\ref{eomchi3}), $D_0$ is the diffusion coefficient
of an isolated Brownian particle, $S(k)$ denotes 
the structure factor, $n$ is the density 
and $M^{\mathrm{irr}}(k;t)$ is the irreducible \cite{irr} memory function.
Moreover, $v_{\mathbf{k}}(\mathbf{k}_1,\mathbf{k}_2) = 
\hat{\mathbf{k}}\cdot\left(c(k_1)\mathbf{k}_1+c(k_2)\mathbf{k}_2\right)$
where $c(k)$ is the Fourier transform of the 
direct correlation function.
The source term at the right-hand-side of Eq. (\ref{eomchi3}), 
$\mathcal{S}_{\mathbf{q}}(\mathbf{k};t;\omega)$, can be expressed in terms of 
the intermediate scattering function and equilibrium correlation functions. 
This term is regular and finite in the $\mathbf{q}\to 0$ and $\omega\to 0$ limits, 
$\lim_{\mathbf{q}\to \mathbf{0}; \omega\to 0} 
\mathcal{S}_{\mathbf{q}}(\mathbf{k};t;\omega) = 
n D_0 S(0) k^2 c(k) F(k;t)$.

As explained below, the re-summation leading to Eqs. (\ref{finalS4om}-\ref{eomchi3}) 
involves an approximation which is similar to the approximation used in
the diagrammatic derivation of the mode-coupling theory \cite{GSdiagram}. 
Thus, to be consistent we replace the full $G$, $F$ and $M^{\mathrm{irr}}$
in Eqs. (\ref{finalS4om}-\ref{eomchi3}) by their approximate forms calculated
using the mode-coupling equations. Specifically, we assume  
the standard mode-coupling relation which 
expresses $M^{\mathrm{irr}}$ in terms of $F$'s, 
$M^{\mathrm{irr}}(k;t) = 
(nD_0/2) \int (d \mathbf{k}'/(2\pi)^3)
v_{\mathbf{k}}^2(\mathbf{k}',\mathbf{k}-\mathbf{k}') 
F(k';t)F(|\mathbf{k}-\mathbf{k}'|;t)$.

At this point we note that evolution operator on the left-hand-side of 
Eq. (\ref{eomchi3}) with 
the mode-coupling expression for $M^{\mathrm{irr}}$  
is exactly the same as the overdamped version of the evolution operator 
derived by BBMR for their three-point susceptibility. Indeed, the only difference 
between Eq. (\ref{eomchi3}) and the overdamped version of the equation
of motion (2) of  Ref. \cite{BBMR} is the source term. Moreover, BBMR emphasized
that all of their qualitative conclusions are independent of the precise
structure of the source term. Thus, we conjecture that our three-point 
functions $\chi^1_{\mathbf{q}}(\mathbf{k};t;\omega)$ 
and $\chi^2_{\mathbf{q}}(\mathbf{k};t;\omega)$ have the same scaling
behavior as $\chi_{\mathbf{q}}(\mathbf{k};t)$ of Ref. \cite{BBMR}. In particular,
these functions  
exhibit a small wave vector divergence upon approaching the
mode-coupling transition. 

Furthermore, we note that Eq. (\ref{finalS4om}) has  
the same general structure as the formula derived by Berthier 
\textit{et al.} (see Eq. (56) of Ref. \cite{JCPa}) \cite{commentd}.
Thus, we are lead to the
same conclusion: the divergence of the four-point dynamic density correlation function
is intimately connected to and in fact driven by the divergence of the three-point
function. 


\textit{Derivation} ---
We consider the time-dependent correlation function
of the projected two-particle density, 
$\left<n_2(1,2;t)n_2^*(3,4)\right>$. 
For a specific set of wave vectors this function is proportional
to $S_4$: in the 
thermodynamic limit we get 
$\left<n_2(\mathbf{k},-\mathbf{k};t)n_2^*(\mathbf{k}-\mathbf{q},\mathbf{k}') \right> = 
n^2 S_4(\mathbf{k};\mathbf{q};t) (2\pi)^3 \delta(\mathbf{k}+\mathbf{q}-\mathbf{k}')$.

Following Ref. \cite{GSdiagram}, 
it is possible to derive a hierarchy of equations
describing the time evolution of $\left<n_2(t) n_2^* \right>$. 
This hierarchy can be rewritten as a hierarchy of integral equations
which can be solved by iteration. The solution can be represented
in terms of diagrams consisting of  
the bare propagator, $G_0(k;t) = \theta(t) \exp(-D_0k^2t/S(k))$,
represented by \includegraphics[scale=.22]{G0a.eps},
and three vertices, the left vertex, 
$\mathcal{V}_{12}$,
represented by \includegraphics[scale=.22]{V12a.eps},
the right vertex, $\mathcal{V}_{21}$, represented by 
\includegraphics[scale=.22]{V21a.eps}, and the four-leg vertex, 
$\mathcal{V}_{22}$, represented by \includegraphics[scale=.2]{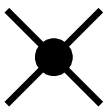}.
The vertices represent renormalized inter-particle
interactions; they involve the direct correlation function and the structure
factor rather than a bare interaction potential 
(see Eqs. (55-57) of Ref. \cite{GSdiagram}).

\begin{figure}
\includegraphics[scale=.22]{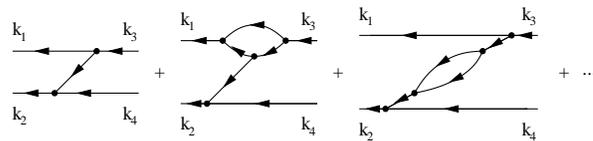}
\caption{The first few diagrams in the diagrammatic series for 
$\left<n_2(\mathbf{k}_1,\mathbf{k}_2;t) 
n_2^*(\mathbf{k}_3,\mathbf{k}_4) \right>^{\mathrm{scn}}$. } 
\label{f:diag2}
\end{figure}

The diagrams contributing to $\left<n_2(t) n_2^*\right>$ separate
into two classes: disconnected and connected ones. 
For $q\neq 0$ only the latter diagrams contribute to 
$S_4(\mathbf{k};\mathbf{q};t)$.
Here, we only consider singly connected diagrams. 
A singly connected diagram is a connected diagram 
which separates into two disconnected components upon cutting a single
$G_0$ bond, with each component containing
at least a single line of bonds from the initial time $0$ to the final time $t$.
The first few such diagrams are shown in Fig. \ref{f:diag2}.

It is convenient to formulate the diagrammatic expansion for the 
singly connected part of $\left<n_2(t) n_2^*\right>$ divided
by $n^2 S(3)S(4)$. For brevity, we will denote this function 
by $\left<n_2(t) n_2^*\right>^{\mathrm{scn}}$. In the diagrams contributing to it, 
we refer to the leftmost and the rightmost bare propagators as the 
left roots and the right roots, respectively, 
and to the other bare propagators as bonds. The roots
are labeled by wave vectors and the bonds are unlabeled. It can be showed that 
$\left<n_2(t) n_2^*\right>^{\mathrm{scn}}$ is the sum of all topologically
different \cite{GSdiagram},  
singly connected diagrams with two left roots, two right roots, $G_0$
bonds, and $\mathcal{V}_{12}$, $\mathcal{V}_{21}$, and $\mathcal{V}_{22}$ 
vertices. To evaluate a 
diagram one assigns wave vectors to bonds and integrates 
over these wave vectors (with a $(2\pi)^{-3}$ factor for each integration),
integrates over all intermediate times, 
and divides the result by a symmetry number \cite{GSdiagram} of the
diagram; diagrams with odd and even numbers of $\mathcal{V}_{22}$
vertices contribute with overall negative and positive sings, respectively.

To re-sum singly connected diagrams we first note that the sum of all the 
``connecting parts'' in diagrams in Fig. \ref{f:diag2} gives the full propagator
$G(q;t)$ \cite{future}. The sums of 
the two remaining parts of singly connected diagrams are related to each other 
by a symmetry operation and thus we have
to evaluate only one of them. 

\begin{figure}
\includegraphics[scale=.22]{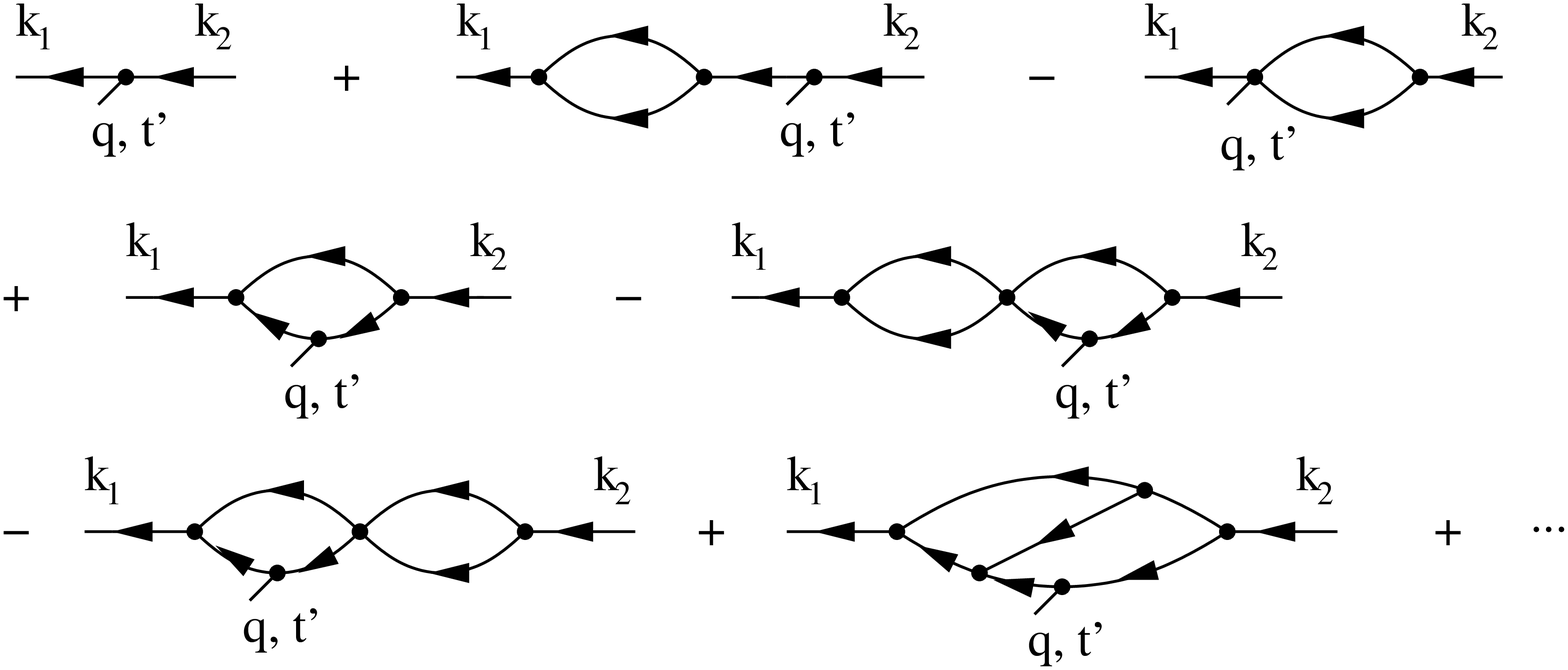}
\caption{The first few diagrams in the diagrammatic expansion for
$X^1_3(\mathbf{k}_1,\mathbf{q};\mathbf{k}_3;t,t')$.
Note that the third, fifth and sixth diagrams, which contain a single 
$\mathcal{V}_{22}$ vertex, contribute with a negative sign.} 
\label{f:diag3}
\end{figure}

We define 
$X^1_3(\mathbf{k}_1,\mathbf{q};\mathbf{k}_3;t,t')$ as the sum of all connected, 
topologically different diagrams with one left root, one right root and
one \textit{left} side root (the left side root originates from a left ``dangling''
end of the right vertex, $\mathcal{V}_{21}$, or from a left ``dangling'' 
end of four-leg vertex, $\mathcal{V}_{22}$). The first few diagrams contributing
to $X^1_3$ are shown in Fig. \ref{f:diag3}.

\begin{figure}
\includegraphics[scale=.22]{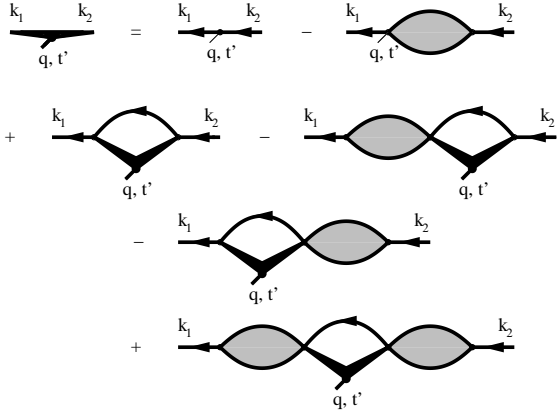}
\caption{Diagrammatic self-consistent equation for function 
$X^1_3(\mathbf{k}_1,\mathbf{q};\mathbf{k}_2;t,t')$.} 
\label{f:diag4}
\end{figure}

The last step in the re-summation is the derivation of a self-consistent equation for 
function $X^1_3$ shown in Fig. \ref{f:diag4}. 
In this figure, \includegraphics[scale=.22]{Xc.eps} represents
function $X^1_3$, a thick bond \includegraphics[scale=.22]{Ga.eps} represents
the full propagator $G(k;t)$ and \includegraphics[scale=.22]{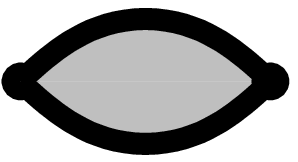} represents
the \textit{reducible} memory matrix \cite{GSdiagram}. 
Here, we will only argue that equation in Fig. \ref{f:diag4} is plausible \cite{future}. 
Re-summation of diagrams similar to the first and second diagrams in
Fig. \ref{f:diag3} results in first diagram at the right-hand-side of 
diagrammatic equation in Fig. \ref{f:diag4}. Re-summation of diagrams similar
to the third diagram in Fig. \ref{f:diag3} results in the second diagram 
at the right-hand-side of Fig. \ref{f:diag4}. Re-summation of
diagrams similar to the fourth diagram in Fig. \ref{f:diag3} 
results in the third diagram at the right-hand-side of Fig. \ref{f:diag4}.
Re-summations of diagrams similar to the fifth and sixth 
diagrams in Fig. \ref{f:diag3} result
in the fourth and fifth diagrams at the right-hand-side of 
Fig. \ref{f:diag4}, respectively. 
Diagrams similar to the seventh diagram in Fig. \ref{f:diag3} are neglected. 
As mentioned above, diagrams neglected in our re-summation 
resemble diagrams which are
neglected in the approximation leading to the mode-coupling expression
for $M^{\mathrm{irr}}$ \cite{GSdiagram}. 

Parenthetically, diagrams resulting from the self-consistent equation showed
in Fig. \ref{f:diag4} could be written as ladder diagrams. However, these
ladder diagrams would be quite unusual in that their beams would correspond
to time running in opposite directions.

Next, we define three-point function  
$\chi^1_{\mathbf{q}}(\mathbf{k};t;t')$ in terms of $X^1_3$: 
$X^1_3(\mathbf{k},\mathbf{q};\mathbf{k}';t,t') = 
\chi^1_{\mathbf{q}}(\mathbf{k};t;t') 
(2\pi)^3 \delta(\mathbf{k}+\mathbf{q}-\mathbf{k}')/S(k')$. We
introduce its frequency dependent version  
$\chi^1_{\mathbf{q}}(\mathbf{k};t;\omega) = \int_0^t dt' e^{i\omega t'} 
\chi^1_{\mathbf{q}}(\mathbf{k};t;t')$. It can be showed that the
self-consistent equation for $X^1_3$ showed in Fig. \ref{f:diag4} 
is equivalent to 
the equation of motion (\ref{eomchi3}) for the frequency dependent function  
$\chi^1_{\mathbf{q}}(\mathbf{k};t;\omega)$. A symmetry relation mentioned
above results in the relation between $\chi^1$ and $\chi^2$ given below
Eq. (\ref{finalS4om}). 
Finally, we express the sum of singly connected diagrams
in terms of $G$, $\chi^1$ and $\chi^2$, and 
we get expression (\ref{finalS4om}) for the part of $S_4$ which exhibits
small wave vector divergence at the mode-coupling transition.

\textit{Conclusions} --- We have showed that the contribution to the 
four-point dynamic density correlation function of a glassy colloidal
suspension due to diagrams which 
separate into two disconnected components upon cutting a single propagator
can be expressed in terms of three-point functions which diverge
at small wave vectors. Thus, the four-point correlation function of a glassy 
colloidal suspension exhibits a small wave vector divergence. In addition, 
we have derived an explicit formula
for the dominant part of the four-point function. 

\textit{Acknowledgments} --- 
I gratefully acknowledge the support of NSF Grant No.~CHE 0517709.



\begin{thebibliography}{99}
\bibitem{Fourier} For brevity, we use the term density for the Fourier 
transform of the density.
\bibitem{commentd} The detailed form
of (\ref{finalS4om}) cannot be compared with results of Refs. \cite{JCPa,JCPb,BerthierS}
since the latter works did not present an explicit formula for the 
four-point function at $q\neq 0$.
\bibitem{future} The details 
will be presented elsewhere.
\end{thebibliography}
\end{document}